\documentclass[twocolumn,prl,showpacs,floatfix]{revtex4}

\usepackage{graphicx}
\usepackage{bm}
\usepackage{color}
\usepackage{amsmath}
\usepackage{amsfonts}
\usepackage{amssymb}

\begin{document}

\preprint{APS/123-QED}

\title{Low-temperature magneto-thermal transport investigation of a Ni-based superconductor BaNi$_2$As$_{2}$: Evidence for fully gapped superconductivity}

\author{N. Kurita$^{1}$} \author{F. Ronning$^1$} \author{Y. Tokiwa$^1$} \author{E. D. Bauer$^1$} \author{A. Subedi$^{2,3}$} \author{D. J. Singh$^2$} \author{J. D. Thompson$^1$} \author{R. Movshovich$^1$}
\affiliation{$^1$Los Alamos National Laboratory, Los Alamos, New Mexico 87545, USA}
\affiliation{$^2$Materials Science and Technology Division, Oak Ridge National Laboratory, Oak Ridge,
Tennessee 37831-6114, USA} \affiliation{$^3$Department of Physics and Astronomy, University of
Tennessee, Knoxville, Tennessee 37996-1200, USA}

\date{\today}

\begin{abstract}
We have performed low-temperature specific heat and thermal conductivity measurements of the Ni-based
superconductor BaNi$_2$As$_{2}$ ($T_\mathrm{c}$\,=\,0.7\,K) in magnetic field. In zero field, thermal
conductivity shows $T$-linear behavior in the normal state and exhibits a BCS-like exponential decrease
below $T_\mathrm{c}$. The field dependence of the residual thermal conductivity extrapolated to zero
temperature is indicative of a fully gapped superconductor. This conclusion is supported by the analysis
of the specific heat data, which are well fit by the BCS temperature dependence from $T_c$ down to the
lowest temperature of 0.1\,K.
\end{abstract}

\pacs{74.70.Dd,74.25.Fy,74.25.Op}
\maketitle

Since the discovery of superconductivity in LaFeAs(O,F)\,\cite{KamiharaJACS2008}, there has been
considerable interest in the oxy-pnictide $R$$T$$Pn$(O,F) and the related structure $A$$T_2$$Pn_2$
($R$\,=\,La, Ce, Sm, Nd, $A$\,=\,Ca, Ba, Sr, Eu, $T$\,=\,Fe, Ni, $Pn$\,=\,P, As) due to (i) their high
superconducting temperature $T_\mathrm{c}$ (up to 55\,K for SmFeAs(O,F)\,\cite{ZARen2008a} and 38\,K for
(Ba,K)Fe$_2$As$_2$\,\cite{Rotter2008b}), (ii) their proximity to magnetism, and (iii) the large variety
in structure and composition that supports superconductivity. Ni-based materials differ from their
Fe-based cousins in that (i) long-range magnetic order has not yet been observed in close proximity to
superconductivity, although similar structural transitions are found, and (ii) $T_\mathrm{c}$ does not
exceed 5\,K in any of the Ni-based systems, although for virtually every structural variant where the Fe
system superconducts, so too does the Ni analog. Substitution of Ni for Fe modifies a number of
properties that may be responsible for suppression of $T_c$. Identifying the commonalities and,
especially, the differences between the compound in the two families, will in turn provide clues into
why $T_c$ in Fe-based superconductors is so high.

One important issue to resolve with any new superconductor is to identify the superconducting order
parameter as that may shed light into the pairing mechanism. Comparing the character of the
superconducting gap in Ni- and Fe-based systems present an ideal opportunity to elucidate the pairing
mechanism responsible for the relatively high $T_\mathrm{c}$ in Fe-based compounds. In the iron-based
oxy-pnictides or $A$Fe$_2$$Pn_2$ compounds, NMR measurements suggest a nodal gap structure\,\cite
{NakaiJPSJ2008,MatanoERL2008}, which contradicts penetration depth\,\cite{Hashimoto2008c}, Andreev
spectroscopy\,\cite{XHChenNature2008}, and angle-resolved photoelectron spectroscopy (ARPES)
measurements\,\cite{DingEPL2008} that indicate a fully gapped multiband superconductor. To our
knowledge, similar measurements that address the character of the superconducting order parameter have
not been done on the Ni
analogs\,\cite{Watanabe2007LaNiPAsO,Mine2008BaNi2P2,FujiiJPCM2008,RonningJPCM2008BaNi2As2,Bauer2008SrNi2As2,Klimczuk2008La3Ni4P4O2,KozhevnikovLaONiBic}.

We present a study of the gap structure of BaNi$_2$As$_{2}$ using low temperature thermal conductivity
and heat capacity. Thermal conductivity, as a bulk probe of low energy delocalized excitations, has
proven to be a powerful tool for identifying nodal\,\cite{dwaveBCS}, $s$-wave\,\cite{dirtySwaveSC},
and multiband superconductors\,\cite{multibandswave}. We chose to study BaNi$_2$As$_{2}$ as a
representative Ni-based superconductor. Bulk superconductivity in this material at 0.7\,K is confirmed
by heat capacity measurements, while, similar to many Fe-pnictides, a first order structural transition
exists at 130\,K\,\cite{RonningJPCM2008BaNi2As2}.

Single crystals of BaNi$_2$As$_{2}$ (ThCr$_2$Si$_2$-type tetragonal structure) were grown in Pb flux as
described in Ref.\,\onlinecite{RonningJPCM2008BaNi2As2}. Thermal conductivity was measured between
40\,mK and 4\,K by a standard one-heater and two-thermometers technique on plate-like crystals with
dimensions of $\sim$\,1\,$\times$\,0.5\,$\times$\,0.1\,mm$^3$ for a heat current
$q$\,$\parallel$\,[100]. Specific heat of several single crystals (6.91\,mg) were measured by a standard heat pulse method. In the thermal conductivity study, two samples from different batches were measured, which are denoted
as $^\#$1 and $^\#$2. Sample $^\#$1 is  the same one used in previous resistivity experiments
(Ref.\,\onlinecite{RonningJPCM2008BaNi2As2}). Magnetic field was applied within $ab$-plane for
$H$$\perp$$q$ ($^\#1$ and $^\#2$) and $H$$\parallel$$q$ ($^\#1$).

\begin{figure}
\begin{center}
\includegraphics[width=0.95\linewidth]{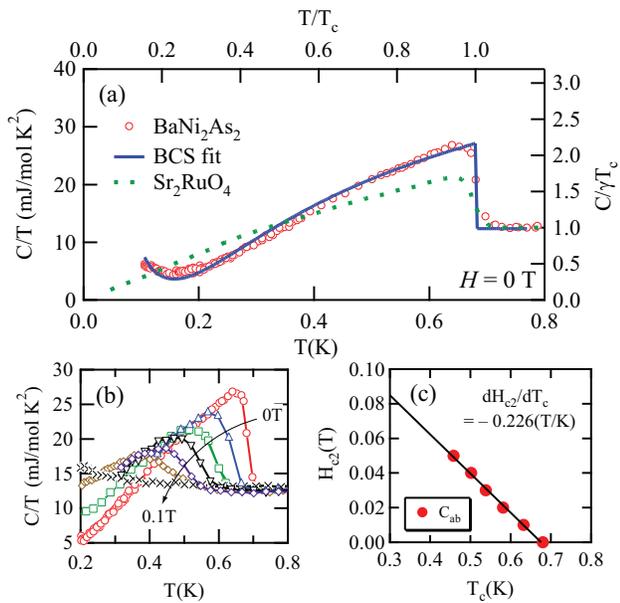}
\end{center}
\caption{(Color online) (a) Zero-field low-temperature specific heat, plotted as $C$/$T$ vs $T$, of BaNi$_2$As$_{2}$ and Sr$_2$RuO$_{4}$\,\cite{Cp_Sr2RuO4} (with $T_\mathrm{c}$ and $\gamma$$T_\mathrm{c}$ scaled to that of BaNi$_2$As$_{2}$) for comparison. The solid curve is a fit to the theoretical expectation for BCS superconductivity including a nuclear quadrupolar Schottky contribution. (b) $C$/$T$ vs $T$ in  fields of 0, 0.01, 0.02, 0.03, 0.04, 0.05 and 0.1\,T (right to left as indicated by the arrow) for $H$$\parallel$${ab}$. (c) Upper critical field $H_\mathrm{c2}$ for $H$$\parallel$${ab}$ as determined by the midpoint of the jump in $C$/$T$. The solid line represents a least-squares fit to the data. } \label{fig1}
\end{figure}

First, we consider the possible superconducting gap symmetry of BaNi$_2$As$_{2}$ based on heat capacity
data. Figure~\ref{fig1}(a) shows specific heat divided by temperature $C$/$T$ as a function of $T$ in
zero field which supports the notion that BaNi$_2$As$_{2}$ is a weak coupling superconductor. The upturn
below 0.2\,K is ascribed to a nuclear quadrupolar Schottky anomaly arising mainly from As. The solid
line is a fit to $C$\,=\,$C_\mathrm{BCS}$\,$+$\,$C_\mathrm{Sch}$, where $C_\mathrm{Sch}$=$A$/$T^2$ and
$C_\mathrm{BCS}$ is the BCS expression:
\begin{eqnarray*}
C_\mathrm{BCS}=t\frac{d}{dt}\int_0^{\infty}\,dy(-\frac{6\gamma\Delta_0}{k_\mathrm{B}\pi^2})[f\ln{f} + (1-f)\ln{(1-f)}].
\end{eqnarray*}
Here $t$\,=\,$T$/$T_\mathrm{c}$, $f$\,=\,1/[$\exp$($E$/$k_\mathrm{B}$$T$)+1],
$E$\,=\,($\epsilon^2$+$\Delta^2$)$^{1/2}$ and $y$\,=\,$\epsilon$/$\Delta_0$, as described in
Ref.\,\onlinecite{Tinkham}. As seen in Fig.~\ref{fig1}(a), the specific heat data can be fit over a wide
temperature range from just above $T_\mathrm{c}$ down to the lowest temperature, yielding a Sommerfeld
coefficient $\gamma$\,=\,12.3\,mJ/mol\,K$^{2}$, $A$\,=\,8.6\,$\times$\,10$^{-3}$\,mJ\,K/mol, equivalent
to the energy splitting of the nuclear quadrupolar levels of As of $\nu_\mathrm{Q}$\,$\approx$\,30\,MHz\,\cite{As_Nuclear}, and $\Delta_0$\,=\,0.0946\,meV\,=\,1.61\,$k_\mathrm{B}$$T_\mathrm{c}$
($T_\mathrm{c}$\,=\,0.68\,K was fixed during the fit). The reduced magnitude of the energy gap
$\Delta_0$, compared with the expected BCS gap
$\Delta_\mathrm{BCS}$\,=\,0.103\,meV\,=\,1.76\,$k_\mathrm{B}$$T_\mathrm{c}$ for a weak coupling BCS
superconductor, is likely due to the fact that our BaNi$_2$As$_{2}$ samples are in the dirty limit as
will be discussed below based on the analysis of thermal conductivity data. In contrast, specific heat
of a well established unconventional superconductor Sr$_2$RuO$_{4}$, also displayed in
Fig.~\ref{fig1}(a), shows a much slower decrease with temperature, expected for nodal
superconductivity\,\cite{Cp_Sr2RuO4}.

Fig.~\ref{fig1}(b) shows the temperature dependence of $C$/$T$ in several fields for
$H$$\parallel$${ab}$ in BaNi$_2$As$_{2}$. As field increases, the jump shifts to low temperature region,
whereas it becomes broader and less detectable above 0.05\,T. The upper critical field $H_\mathrm{c2}$,
determined by the midpoint of the jump at each field, is displayed in Fig.~\ref{fig1}(c). The solid line
is a least-squares fit to the $H_{c2}$ data, yielding an initial slope
d$H_\mathrm{c2}$/d$T_\mathrm{c}$ =$-$0.226\,T/K, which results in  a zero-temperature upper critical
field $H_\mathrm{c2}$(0)\,=\,0.11\,T, using the estimate
$H_\mathrm{c2}$(0)\,=\,$-$0.7\,$T_\mathrm{c}$d$H_\mathrm{c2}$/d$T_\mathrm{c}$\,\cite{WHH1966}. This
yields a Ginzburg-Landau coherence length $\xi$\,=\,550\,$\mathrm{\AA}$ from the relationship
$\xi$\,=\,($\mathrm{\Phi}_0$/2$\mathrm{\pi}$$H_\mathrm{c2}$(0))$^{1/2}$, where
$\Phi_0$\,=\,2.07\,$\times$\,10$^{-7}$\,Oe\,cm$^2$ is the flux quantum. An estimate of the electronic
mean free path $l_\mathrm{e}$ is obtained from the normal state thermal conductivity using
$\kappa$/$T$\,=\,1/3$\gamma$$v_\mathrm{F}$$l_\mathrm{e}$, with a Fermi velocity
$v_\mathrm{F}$\,=\,2.96\,$\times$\,10$^5$\,m/s which was calculated by methods described in
Ref.\,\onlinecite{Sudedi2008DensityBaNi2As2}. We find $l_\mathrm{e}$\,$\approx$\,70\,$\mathrm{\AA}$ and
$l_\mathrm{e}$/$\xi$\,=\,0.13, which places BaNi$_2$As$_{2}$ in the dirty limit. The field dependence of
the residual linear term $\gamma_0$$T$ of the heat capacity can be used to identify whether or not the
superconducting order parameter has nodes. However, in the low temperature region, especially below
0.2\,K, the addenda contribution, as well as the nuclear term, grow with increasing field, making an
accurate estimate of $\gamma_0(H)$ difficult. In this regard, thermal conductivity, which is not
subject to these effects, is ideally suited for directly determining the low-temperature behavior.

\begin{figure}
\begin{center}
\includegraphics[width=0.95\linewidth]{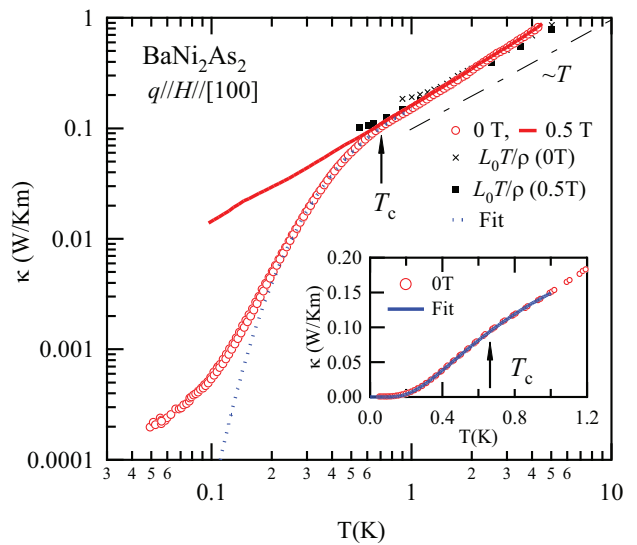}
\end{center}
\caption{(Color online) Temperature dependence of the thermal conductivity $\kappa$($T$) in BaNi$_2$As$_{2}$ for heat current $q$$\parallel$[100], in zero field and in the normal state (0.5\,T) for $H$$\parallel$$ab$. Arrows indicate $T_\mathrm{c}$ determined from the specific heat measurement. Solid square and cross symbol represent the electronic thermal conductivity $\kappa$$_e$\,=\,$L_0$$T$/$\rho$ with $L_0$\,=\,2.45\,$\times$\,10$^{-8}$\,W$\Omega$/K$^2$ for 0 and 0.5\,T, respectively, derived from resistivity data using the Wiedemann-Franz law. Dotted line in main figure and solid line in the inset show a fit to $\kappa$($T$) with a BCS curve defined as $\kappa$\,=\,$C$exp(${-a}$$T_\mathrm{c}$/$T$).} \label{fig2}
\end{figure}

Fig.~\ref{fig2} shows the temperature dependence of thermal conductivity $\kappa$($T$) of
BaNi$_2$As$_{2}$ in zero field and 0.5\,T ($\gg$$H_\mathrm{c2}$) for heat current $q$$\parallel$[100]. In
zero field and in the normal state above $T_\mathrm{c}$\,=\,0.7\,K, $\kappa$($T$) exhibits an
approximately $T$-linear variation, which continues to the lowest temperature in a field of 0.5\,T.
Using the Wiedemann-Franz law, we estimate an electronic thermal conductivity in the normal state
$\kappa$$_e$\,=\,$L_0$$T$/$\rho$ in zero-field and 0.5\,T where the Lorenz number
$L_0$\,=\,2.44\,$\times$\,10$^{-8}$\,W$\mathrm{\Omega}$/K$^2$. As seen in Fig.~\ref{fig2},
$\kappa$$_e$($T$) is very close to $\kappa$($T$) in the normal state, suggesting that heat transport in
the normal state is dominated by the electrons. In the superconducting state, $\kappa$($T$) follows an
exponential form ($\sim$\,exp($-$$a$$T_\mathrm{c}$/$T$) with $a$\,=\,1.34) down to 0.2\,K as shown in
the inset of Fig.~\ref{fig2}. This evolution of $\kappa$($T$) in BaNi$_2$As$_{2}$ is quite similar to
the conventional superconductors tin, aluminum, and zinc with $T_\mathrm{c}$\,=\,3.7, 1.2 and 0.84\,K,
respectively\,\cite{Berman}. The thermal conductivity for these three elements is also described by
$\kappa$($T$)\,$\sim$\,exp($-$$a$$T_\mathrm{c}$/$T$) with $a$\,=\,1.3\,-\,1.5 below $T_\mathrm{c}$. In
fact, the thermal conductivity based on BCS superconductivity for weak coupling\,\cite{VRT_kappaBCS} is
well-described by the formula exp($-$$a$$T_\mathrm{c}$/$T$) with $a$\,=\,1.47. Thus, the $a$-value of
1.34 for BaNi$_2$As$_{2}$ is in good agreement with that for conventional BCS superconductors. Below
0.2\,K, $\kappa$($T$) does not fall as rapidly as expected for an exponential dependence  and reaches
2\,$\times$\,10$^{-4}$\,W/K\,m at 50\,mK. Similar deviations from an exponential dependence were
observed in a number of conventional superconductors, including tin and aluminum\,\cite{Berman}, and are usually attributed to phonon contribution. When
most of the normal quasiparticles are frozen out, the intrinsic electronic thermal conductivity is too
low to make an appreciable contribution. In fact, an upper limit of the phonon thermal conductivity
estimated via
$\kappa_\mathrm{ph}$\,=\,$\frac{1}{3}$$\beta$$T^3$$\langle$$v$$\rangle$$l_\mathrm{ph}$\,=\,2.0\,$\times$\,10$^{-4}$\,W/K\,m,
using the phonon specific coefficient $\beta$\,=\,18.7\,J/K$^{4}$\,m$^3$, mean phonon velocity
$\langle$$v$$\rangle$\,=\,1860\,m/s\,\cite{RonningJPCM2008BaNi2As2} and the phonon mean free path
$l_\mathrm{ph}$\,=\,1.38\,$\times$\,10$^{-4}$\,m\,\cite{meanfreepath}, is equal to the experimental
value 2\,$\times$\,10$^{-4}$\,W/K\,m at 50\,mK in BaNi$_2$As$_{2}$. Although we cannot rule out other
origins of the low temperature tail, agreement between the estimated and the measured values at 50\,mK
suggests a phonon origin of the low temperature thermal conductivity in BaNi$_2$As$_{2}$. Note that
nuclear quadrupolar moments, while leading to a Schottky term dominating specific heat below 200 mK, do
not contribute to thermal conductivity.

\begin{figure}
\begin{center}
\includegraphics[width=0.95\linewidth]{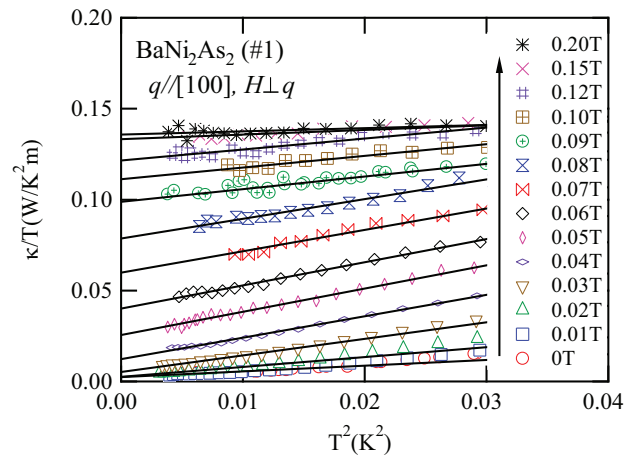}
\end{center}
\caption{(Color online) Low-temperature dependence of  $\kappa$/$T$ vs $T^2$ for
$q$$\parallel$[100] in several fields for $H$$\perp$$q$. The solid lines are fits to the data in
the temperature range by $\kappa$/$T$\,=\,$\kappa_0$/$T$\,+\,$b$$T^2$} \label{fig3}
\end{figure}

\begin{figure}
\begin{center}
\includegraphics[width=0.95\linewidth]{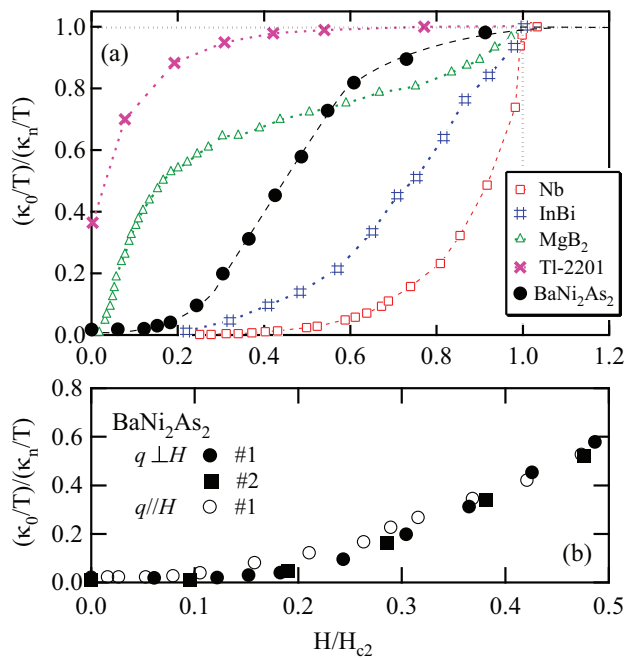}
\end{center}
\caption{(Color online) (a) Residual thermal conductivity $\kappa_0$/$T$ of BaNi$_2$As$_{2}$, normalized
by the normal state value $\kappa_\mathrm{n}$/$T$ above $H_\mathrm{c2}$, as a function of
$H$/$H_\mathrm{c2}$ for $H$$\perp$$q$ ($^\#1$). A small contribution from  the Pb flux has been
subtracted. For comparison, data for several superconductors with different superconducting gap characteristics are shown: Nb (clean, fully gapped $s$-wave)\,\cite{Nb}, InBi (dirty, fully gapped $s$-wave)\,\cite{dirtySwaveSC}, MgB$_2$ (multi-band gap)\,\cite{multibandswave}, Tl-2201 ($d$-wave with line nodes)\,\cite{dwaveBCS}. The dotted lines are guides to the eyes. (b)
($\kappa_0$/$T$)/($\kappa_\mathrm{n}$/$T$) vs $H$/$H_\mathrm{c2}$ of BaNi$_2$As$_{2}$ for $H$$\perp$$q$
($^\#1$,$^\#2$) and $H$$\parallel$$q$ ($^\#1$).} \label{fig4}
\end{figure}

The low-temperature magnetic field dependence of $\kappa$ is instructive for determining the gap
structure. Fig.~\ref{fig3} shows the low-temperature expansion of $\kappa$/$T$ vs $T^2$ of
BaNi$_2$As$_{2}$ for a heat current $q$$\parallel$[100] in several fields for $H$$\perp$$q$. The
straight lines are fits to $\kappa$/$T$\,=\,$\kappa_0$/$T$\,+\,$b$\,$T^2$, where $\kappa_0$/$T$ is the
residual term extrapolated to $T$\,=\,0\,K at each field. With increasing field, $\kappa_0$/$T$\ rapidly
increases above 0.02\,T and saturates above 0.15\,T. This rise could only be attributed to the electron
contribution, as the phonon thermal conductivity may only go down in magnetic field due to additional
scattering from vortices in the mixed state. The field dependence of the residual thermal conductivity
of BaNi$_2$As$_{2}$ is shown in Fig.~\ref{fig4}(a) as ($\kappa_0$/$T$)/($\kappa_\mathrm{n}$/$T$) vs
$H$/$H_\mathrm{c2}$ for $H$$\perp$$q$ ($^\#1$). From the thermal conductivity data,
$H_\mathrm{c2}$\,=\,0.16\,T is determined as the field where $\kappa_0$/$T$ becomes independent of the
magnetic field. $\kappa_\mathrm{n}$/$T$ at 0.2\,T is used as the normal state value. Data for several
superconductors with different gap structures are also shown for comparison. The simplest test of
$s$-wave superconductivity is whether the residual thermal conductivity exhibits a concave variation
with field as vortices begin to penetrate the sample at low fields. The exponential-like concave
behavior is understood as follows: when vortices first enter the sample at $H$$\ge$$H_{c1}$,
quasiparticles that contribute to thermal conductivity are mostly localized around the vortex cores.
Therefore, at low field, those quasiparticles cannot carry heat until the intervortex spacing is
decreased sufficiently that states created in the vortices begin to overlap, become delocalized, and
hence, transport heat, as illustrated by the clean $s$-wave superconductor Nb\cite{Nb}). In contrast,
the presence of nodal quasiparticles leads to a convex field dependence as a consequence of the Volovik
effect\,\cite{Volovik}, as illustrated by Tl-2201,\cite{dwaveBCS}. As seen in Fig.~\ref{fig4}(b),
($\kappa_0$/$T$)/($\kappa_\mathrm{n}$/$T$) exhibits concave feature independent of heat current
direction with respect to magnetic field, and independent of sample difference. Consequently,
\textit{the concave field-dependence of the thermal conductivity implies that BaNi$_2$As$_{2}$ is a
fully gapped superconductor}.

The field dependence of $\kappa$($H$) for BaNi$_2$As$_{2}$ more closely resembles the dirty $s$-wave
superconductors such as InBi, than the clean case (i.e., Nb), which is consistent with our estimate of
$l_\mathrm{e}$/$\xi$\,$\ll$\,1. In addition, the data appear to display a shoulder-like anomaly at
0.7\,$H_\mathrm{c2}$, suggesting an additional energy scale at that field. This may reflect a spread in
$H$$_\mathrm{c2}$ as reflected by the width of the heat capacity anomaly in field. On the other hand,
the shoulder might be due to multiband character of superconductivity in BaNi$_2$As$_{2}$, although the
anomaly in thermal conductivity is much less dramatic than in MgB$_2$. And there is no obvious evidence for multiband physics in the heat capacity data of Fig.~\ref{fig1}. However, interband scattering
would be expected to wipe out multiband effects in our relatively dirty material. Additional studies on
cleaner crystals would help to resolve this issue.

For the FeAs-based superconductors, it has been argued that an unconventional superconducting state may
still give rise to a fully gapped excitation spectrum if the nodal planes, where the phase of the order
parameter changes sign, do not intersect the Fermi surface\,\cite{Seo2008c}. Density functional
calculations for BaNi$_2$As$_{2}$, however, find a much more complicated Fermi surface for which it is
difficult to imagine finding a nodal plane which does not intersect the Fermi
surface\,\cite{Sudedi2008DensityBaNi2As2}. Thus, the fully gapped Fermi surface in BaNi$_2$As$_2$ is due
to the superconducting order parameter that does not change sign. If the superconducting order parameter
in FeAs systems does indeed change sign, the pairing mechanism in the FeAs and NiAs systems is likely to
be of different origin, as argued by band structure
calculations\,\cite{Sudedi2008DensityBaNi2As2,Sudedi2008DensityFeAs}.

In conclusion, we have performed magneto-thermal transport experiments to elucidate the superconducting
gap symmetry of BaNi$_2$As$_{2}$. The following results indicate that BaNi$_2$As$_{2}$ is a dirty fully
gapped superconductor: (i) specific heat data are well-fit by the BCS formula, (ii) the coherence length
is much larger than the electronic mean free path, and, (iii) the field dependence of the residual
thermal conductivity  is consistent with a dirty fully gapped superconductivity.

We would like to thank I. Vekhter M. Graf, S.-H Baek and H. Sakai for useful discussions. Work at Los Alamos National
Laboratory was performed under the auspices of the US Department of Energy. Work at Oak Ridge was
supported by the DOE, Division of Materials Sciences and Engineering.


\begin{thebibliography}{25}
\expandafter\ifx\csname natexlab\endcsname\relax\def\natexlab#1{#1}\fi \expandafter\ifx\csname
bibnamefont\endcsname\relax
  \def\bibnamefont#1{#1}\fi
\expandafter\ifx\csname bibfnamefont\endcsname\relax
  \def\bibfnamefont#1{#1}\fi
\expandafter\ifx\csname citenamefont\endcsname\relax
  \def\citenamefont#1{#1}\fi
\expandafter\ifx\csname url\endcsname\relax
  \def\url#1{\texttt{#1}}\fi
\expandafter\ifx\csname urlprefix\endcsname\relax\def\urlprefix{URL }\fi
\providecommand{\bibinfo}[2]{#2} \providecommand{\eprint}[2][]{\url{#2}}




\bibitem[{\citenamefont{Kamihara}(2008)}]{KamiharaJACS2008} \bibinfo{author}{\bibfnamefont{Y.}
    \bibnamefont{Kamihara {\em et al.}}},
  \bibinfo{journal}{J. Am. Chem. Soc.} \textbf{\bibinfo{volume}{130}},
  \bibinfo{pages}{3296} (\bibinfo{year}{2008}).





\bibitem[{\citenamefont{}(2008)}]{ZARen2008a} \bibinfo{author}{\bibfnamefont{Z.-A.} \bibnamefont {Ren
    {\em et al.}}},
  \bibinfo{journal}{Chin. Phys. Lett.} \textbf{\bibinfo{volume}{25}},
  \bibinfo{pages}{2215} (\bibinfo{year}{2008}).




\bibitem[{\citenamefont{Rotter}(2008)}]{Rotter2008b} \bibinfo{author}{\bibfnamefont{M.}
    \bibnamefont{Rotter {\em et al.}}},
  \bibinfo{journal}{Phys. Rev. Lett.} \textbf{\bibinfo{volume}{101}},
  \bibinfo{pages}{107006} (\bibinfo{year}{2008}).




\bibitem[{\citenamefont{Kamihara}(2008)}]{NakaiJPSJ2008} \bibinfo{author}{\bibfnamefont{Y}
    \bibnamefont{Nakai {\em et al.}}},
  \bibinfo{journal}{J. Phys. Soc. Jpn.} \textbf{\bibinfo{volume}{77}},
  \bibinfo{pages}{073701} (\bibinfo{year}{2008}).





\bibitem[{\citenamefont{Kamihara}(2008)}]{MatanoERL2008} \bibinfo{author}{\bibfnamefont{K.}
    \bibnamefont{Matano {\em et al.}}},
  \bibinfo{journal}{Europhys. Lett.} \textbf{\bibinfo{volume}{83}},
  \bibinfo{pages}{57001} (\bibinfo{year}{2008}).





\bibitem[{\citenamefont{}(2008)}]{Hashimoto2008c} \bibinfo{author}{\bibfnamefont{K.}
    \bibnamefont{Hashimoto {\em et al.}}},
  \bibinfo{journal}{Phys. Rev. Lett.} \textbf{\bibinfo{volume}{102}},
  \bibinfo{pages}{017002} (\bibinfo{year}{2009}).





\bibitem[{\citenamefont{}(2008)}]{XHChenNature2008} \bibinfo{author}{\bibfnamefont{X.H.}
    \bibnamefont{Chen {\em et al.}}},
  \bibinfo{journal}{Nature} \textbf{\bibinfo{volume}{453}},
  \bibinfo{pages}{761} (\bibinfo{year}{2008}).




\bibitem[{\citenamefont{}(2008)}]{DingEPL2008} \bibinfo{author}{\bibfnamefont{H.} \bibnamefont{Ding {\em
    et al.}}},
  \bibinfo{journal}{Europhys. Lett.} \textbf{\bibinfo{volume}{83}},
  \bibinfo{pages}{47001} (\bibinfo{year}{2008}).






\bibitem[{\citenamefont{}(2007)}]{Watanabe2007LaNiPAsO} \bibinfo{author}{\bibfnamefont{T.}
    \bibnamefont{Watanabe {\em et al.}}},
  \bibinfo{journal}{Inorg. Chem.} \textbf{\bibinfo{volume}{46}},
  \bibinfo{pages}{7719} (\bibinfo{year}{2007}).
\bibinfo{author}{\bibfnamefont{T.} \bibnamefont{Watanabe {\em et al.}}},
  \bibinfo{journal}{J. Solid State Chem.} \textbf{\bibinfo{volume}{181}},
  \bibinfo{pages}{2117} (\bibinfo{year}{2008}).




\bibitem[{\citenamefont{}(2008)}]{Mine2008BaNi2P2} \bibinfo{author}{\bibfnamefont{T.} \bibnamefont{Mine
    {\em et al.}}},
  \bibinfo{journal}{Solid State Commun.} \textbf{\bibinfo{volume}{147}},
  \bibinfo{pages}{111} (\bibinfo{year}{2008}).



\bibitem[{\citenamefont{}(2008)}]{FujiiJPCM2008} \bibinfo{author}{\bibfnamefont{H.} \bibnamefont{Fujii}}
    and \bibinfo{author}{\bibfnamefont{S.} \bibnamefont{Kasahara}},
  \bibinfo{journal}{J. Phys.: Condens. Matter} \textbf{\bibinfo{volume}{20}},
  \bibinfo{pages}{075202} (\bibinfo{year}{2008}).




\bibitem[{\citenamefont{}(2008)}]{RonningJPCM2008BaNi2As2} \bibinfo{author}{\bibfnamefont{F.}
    \bibnamefont{Ronning {\em et al.}}},
  \bibinfo{journal}{J. Phys.: Condens. Matter} \textbf{\bibinfo{volume}{20}},
  \bibinfo{pages}{342203} (\bibinfo{year}{2008}).




\bibitem[{\citenamefont{}(2008)}]{Bauer2008SrNi2As2} \bibinfo{author}{\bibfnamefont{E.D.}
    \bibnamefont{Bauer {\em et al.}}},
  \bibinfo{journal}{Phys. Rev. B} \textbf{\bibinfo{volume}{78}},
  \bibinfo{pages}{172504} (\bibinfo{year}{2008}).




\bibitem[{\citenamefont{}(2008)}]{Klimczuk2008La3Ni4P4O2} \bibinfo{author}{\bibfnamefont{T.}
    \bibnamefont{Klimczuk {\em et al.}}},
  \bibinfo{journal}{Phys. Rev. B} \textbf{\bibinfo{volume}{79}},
  \bibinfo{pages}{012505} (\bibinfo{year}{2009}).




\bibitem[{\citenamefont{}(2008)}]{KozhevnikovLaONiBic} \bibinfo{author}{\bibfnamefont{V.L.}
    \bibnamefont{Kozhevnikov {\em et al.}}},
  \bibinfo{journal}{JETP. Lett.} \textbf{\bibinfo{volume}{87}},
  \bibinfo{pages}{747} (\bibinfo{year}{2008}).





\bibitem[{\citenamefont{}(2008)}]{dwaveBCS} \bibinfo{author}{\bibfnamefont{C.} \bibnamefont{Proust {\em
    et al.}}},
  \bibinfo{journal}{Phys. Rev. Lett.} \textbf{\bibinfo{volume}{89}},
  \bibinfo{pages}{147003} (\bibinfo{year}{2002}).
\bibinfo{author}{\bibfnamefont{M.} \bibnamefont{Suzuki {\em et al.}}},
  \bibinfo{journal}{Phys. Rev. Lett.} \textbf{\bibinfo{volume}{88}},
  \bibinfo{pages}{227004} (\bibinfo{year}{2002}).




\bibitem[{\citenamefont{}(2008)}]{dirtySwaveSC} \bibinfo{author}{\bibfnamefont{J.} \bibnamefont{Willis}}
    and \bibinfo{author}{\bibfnamefont{D.} \bibnamefont{Ginsberg}},
 \bibinfo{journal}{Phys. Rev. B} \textbf{\bibinfo{volume}{14}},
  \bibinfo{pages}{1916} (\bibinfo{year}{1976}).
\bibinfo{author}{\bibfnamefont{M.} \bibnamefont{Sutherland {\em et al.}}},
  \bibinfo{journal}{Phys. Rev. Lett.} \textbf{\bibinfo{volume}{98}},
  \bibinfo{pages}{067003} (\bibinfo{year}{2007}).





 \bibitem[{\citenamefont{}(2008)}]{multibandswave} \bibinfo{author}{\bibfnamefont{A.V.}
    \bibnamefont{Sologubenko {\em et al.}}},
  \bibinfo{journal}{Phys. Rev. B} \textbf{\bibinfo{volume}{66}},
  \bibinfo{pages}{014504} (\bibinfo{year}{2002}).
\bibinfo{author}{\bibfnamefont{E.} \bibnamefont{Boaknin {\em et al.}}},
  \bibinfo{journal}{Phys. Rev. Lett.} \textbf{\bibinfo{volume}{90}},
  \bibinfo{pages}{117003} (\bibinfo{year}{2003}).



\bibitem[{\citenamefont{}(1966)}]{Tinkham} \bibinfo{author}{\bibfnamefont{See, for example, M.}
    \bibnamefont{Tinkham}},
  \bibinfo{journal}{{\em Introduction to Superconductivity} (McGraw-Hill, New York)}\textbf{\bibinfo{volume}{}}, \bibinfo{pages}{} (\bibinfo{year}{1975}).







\bibitem[{\citenamefont{}(1966)}]{As_Nuclear}
    \bibinfo{author}{\bibfnamefont{$\nu_\mathrm{Q}\approx$\,30\,MHz is
    comparable to values of $\nu_\mathrm{Q}$ for other compounds of this family. For example, $\nu_\mathrm{Q}$\,$=$\,13.93\,MHz for
    CaFe$_2$As$_2$, S.-H.} \bibnamefont{Baek {\em et al.}} arxiv:0808.0744v3. Direct measurements of
    $\nu_\mathrm{Q}$
    in BaNi$_2$As$_2$ via NQR will be useful.}






\bibitem[{\citenamefont{}(1966)}]{Cp_Sr2RuO4} \bibinfo{author}{\bibfnamefont{S.} \bibnamefont{Nishizaki
    {\em et al.}}},
  \bibinfo{journal}{J. Low Temp. Phys.} \textbf{\bibinfo{volume}{117}},
  \bibinfo{pages}{1581} (\bibinfo{year}{1999}).







\bibitem[{\citenamefont{}(1966)}]{WHH1966} \bibinfo{author}{\bibfnamefont{N.R.} \bibnamefont{Werthamer
    {\em et al.}}},
  \bibinfo{journal}{Phys. Rev.} \textbf{\bibinfo{volume}{147}},
  \bibinfo{pages}{295} (\bibinfo{year}{1966}).




\bibitem[{\citenamefont{}(1966)}]{Sudedi2008DensityBaNi2As2} \bibinfo{author}{\bibfnamefont{A.}
    \bibnamefont{Subedi}} and \bibinfo{author}{\bibfnamefont{D.J.} \bibnamefont{Singh}},
 \bibinfo{journal}{Phys. Rev. B} \textbf{\bibinfo{volume}{78}},
  \bibinfo{pages}{132511} (\bibinfo{year}{2008}).
















\bibitem[{\citenamefont{}(1966)}]{Berman} \bibinfo{author}{\bibfnamefont{See, for example, R.}
    \bibnamefont{Berman}},
  \bibinfo{journal}{{\em Thermal conduction in Solids} (Oxford Univ. Press, Oxford),} \textbf{\bibinfo{volume}{}}\bibinfo{pages}{}(\bibinfo{year}{1976}) and references therein.



\bibitem[{\citenamefont{}(1966)}]{VRT_kappaBCS} \bibinfo{author}{\bibfnamefont{J.} \bibnamefont{Bardeen
    {\em et al.}}},
  \bibinfo{journal}{Phys. Rev.} \textbf{\bibinfo{volume}{113}},
  \bibinfo{pages}{982} (\bibinfo{year}{1959}).


\bibitem[{\citenamefont{}(1966)}]{meanfreepath} \bibinfo{author}{\bibfnamefont{} \bibnamefont{}}
  \bibinfo{journal}{$\Lambda_0$\,=\,$\frac{2}{\pi}$$\sqrt{ab}$, where $a$\,=\,73\,$\mathrm{\mu}$m and $b$\,=\,640\,$\mathrm{\mu}$m are sample cross section dimensions. A nearly order of magnitude difference between $a$ and $b$ can lead to an overestimate of $l_\mathrm{ph}$. See, \bibinfo{author}{\bibfnamefont{M.P.} \bibnamefont{Zaitlin {\em et al.}}},
  \bibinfo{journal}{Phys. Rev.} \textbf{\bibinfo{volume}{B 12}},
  \bibinfo{pages}{4487} (\bibinfo{year}{1975}).}



\bibitem[{\citenamefont{}(1983)}]{Nb} \bibinfo{author}{\bibfnamefont{J.} \bibnamefont{Lowell}} and
    \bibinfo{author}{\bibfnamefont{J.B.} \bibnamefont{Sousa}},
  \bibinfo{journal}{J. Low. Temp. Phys.} \textbf{\bibinfo{volume}{3}},
  \bibinfo{pages}{65} (\bibinfo{year}{1970}).

\bibitem[{\citenamefont{}(1966)}]{Volovik} \bibinfo{author}{\bibfnamefont{G.E.} \bibnamefont{Volovik}},
  \bibinfo{journal}{JETP Lett.} \textbf{\bibinfo{volume}{58}},
  \bibinfo{pages}{469} (\bibinfo{year}{1993}).





\bibitem[{\citenamefont{}(1966)}]{Seo2008c} \bibinfo{author}{\bibfnamefont{e.g. K.} \bibnamefont{Seo
    {\em et al.}}},
\bibinfo{journal}{Phys. Rev. Lett.} \textbf{\bibinfo{volume}{101}},
  \bibinfo{pages}{206404} (\bibinfo{year}{2008}).








\bibitem[{\citenamefont{}(1966)}]{Sudedi2008DensityFeAs} \bibinfo{author}{\bibfnamefont{A.}
    \bibnamefont{Subedi {\em et al.}}},
 \bibinfo{journal}{Phys. Rev. B} \textbf{\bibinfo{volume}{78}},
  \bibinfo{pages}{134514} (\bibinfo{year}{2008}).



\end{thebibliography}
\end{document}